\documentclass[conference]{IEEEtran}
\IEEEoverridecommandlockouts
\usepackage{amsmath,amssymb,amsfonts}

\usepackage{graphicx}
\usepackage{textcomp}
\usepackage{xcolor}
\usepackage{url}
\usepackage[numbers,sort]{natbib}
\usepackage[noend]{algpseudocode}
\usepackage{algorithm,algorithmicx}
\newcommand*\Let[2]{\State #1 $\gets$ #2}
\usepackage{etoolbox,xspace}
\algrenewcommand\algorithmicrequire{\textbf{Input:}}
\algrenewcommand\algorithmicensure{\textbf{Postcondition:}}
\usepackage{comment}
\usepackage{caption}
\usepackage{subcaption}
\usepackage{diagbox}
\def\BibTeX{{\rm B\kern-.05em{\sc i\kern-.025em b}\kern-.08em
    T\kern-.1667em\lower.7ex\hbox{E}\kern-.125emX}}

\begin{document}

\title{Multi-View Structural Graph Summaries}

\author{\IEEEauthorblockN{Jonatan Frank, Andor Diera}
\IEEEauthorblockA{\textit{Data Science and Big Data Analytics} \\
Ulm University, Germany \\
firstname.lastname@uni-ulm.de}
\and
\IEEEauthorblockN{David Richerby}
\IEEEauthorblockA{
University of Essex, Colchester, UK \\
david.richerby@essex.ac.uk}
\and
\IEEEauthorblockN{Ansgar Scherp}
\IEEEauthorblockA{\textit{Data Science and Big Data Analytics} \\
Ulm University, Germany \\
ansgar.scherp@uni-ulm.de}
}

\def\news{INC 2023 } 
\def\code{CSS3V } 

\maketitle

\begin{abstract}
A structural graph summary is a small graph representation that preserves structural information necessary for a given task. The summary is used instead of the original graph to complete the task faster. We introduce multi-view structural graph summaries and propose an algorithm for merging two summaries. We conduct a theoretical analysis of our algorithm. We run experiments on three datasets, contributing two new ones. The datasets are of different domains (web graph, source code, and news) and sizes. The interpretation of multi-view depends on the domain: pay-level domains on the web, control vs.\@ data flow of the code, and the output of different news broadcasters. We experiment with three graph summary models: attribute collection, class collection, and their combination. We observe that merging two structural summaries has an upper bound of quadratic complexity; but under reasonable assumptions, it has linear-time worst-case complexity. The running time of merging has a strong linear correlation with the number of edges in the two summaries. Therefore, the experiments support the assumption that the upper bound of quadratic complexity is not tight and that linear complexity is possible. Furthermore, our experiments show that always merging the two smallest summaries by the number of edges is the most efficient strategy for merging multiple structural summaries.
The source code and additional resources are available at \url{https://github.com/jofranky/Multi-View-Structural-Graph-Summaries}.
\end{abstract}

\section{Introduction}
\label{sec:introduction}
Graph summarization aims to generate a small representation $S$ of an input graph $G$, which preserves structural information necessary for a given task~\cite{blumericherbyscherp-tcs-2021}. 
$S$~is called the summary of $G$ and is used for various tasks.
For example, summaries are used to speed up web data search~\cite{DBLP:journals/ws/KonrathGSS12}, for estimating cardinalities in database joins~\cite{DBLP:conf/icde/NeumannM11}, data visualization~\cite{DBLP:journals/vldb/GoasdoueGM20}, 
and graph distillation in machine learning~\cite{gupta2023mirage}.

To the best of our knowledge, we are the first to consider multi-view graphs in the context of graph summarization.
We consider a set $ \mathcal{G} = \{ G_1, \ldots, G_n \} $ of graphs to be multiple views of the same underlying graph.
The interpretation of multi-view graphs depends on the domains and the sources of the graphs. 
For example, in the \news news dataset, a view is the output of a broadcaster, and in the \code dataset, it depends on what aspect of a code is considered, e.\,g., the data flow or the control flow.
Each graph $G_i\in \mathcal{G}$ provides either a different view of the same topic (one domain) or is about a different topic (many domains). 
They also have either the same source (single source) or different sources (multiple sources). 
The respective summaries of a multi-view graph $\mathcal{G}$ are $ \mathcal{S} = \{ S_1, \dots, S_n \}$ and provide smaller representations of the graphs. Merging them into a summary of~$\mathcal{G}$ results in a smaller summary that contains less redundancy. 
Hence, the merged summary enables tasks to be completed faster than the original summaries.

It is natural to merge the different views by taking their union.
However, the same approach does not apply to summaries when we have summaries of the individual views $G_1, \dots, G_n$ and wish to create a summary of the merged graph $G_1\cup\dots\cup G_n$.
The union of the summaries results in inconsistencies, including some vertices being mapped to the wrong summary vertex and the merged summary containing incorrect information.
This occurs, for example, when the views provide different information about the same vertices.

In our work, we introduce multi-view structural graph summarization and an algorithm for merging summaries, independent of the summary model.
For our experiments, we use three datasets: 
the Billion Triple Challenge 2019 dataset (BTC 2019)~\cite{DBLP:conf/semweb/HerreraHK19}, a web graph of 2~billion edges; 
our multi-view source code dataset of $330,000$ R~programs, the CRAN Social Science Three Views dataset (CSS3V);
and our International News Coverage 2023 dataset (INC 2023), created from $450$ real-world news articles by GPT-4~\cite{openai2024gpt4}.
These datasets are from different domains (web graph, source code, and news) and are in the form of Resource Description Framework (RDF) graphs~\cite{w3c-rdf}.
We use attribute collection (AC), class collection (CC),  and attribute class collection (ACC) as summary models~\cite{DBLP:conf/dexaw/CampinasPCDT12}. 
ACC is a combination of AC and CC. 
With ACC, we can investigate how using a combined summary model affects merging two summary graphs, compared to merging summaries based on AC or CC alone.

Our results show that merging two structural summaries has linear worst-case complexity.
This is under the assumption that all necessary information is saved in advance and that hash tables have constant-sized keys.  
In our experiments, pairwise merge times have a strong linear correlation with the number of edges in the summaries. 
Our experiments also show that the most time- and memory-efficient strategy for merging multiple structural summaries is to merge in increasing order of size (number of edges).
The contributions of our work are:

\begin{itemize}

\item We introduce an algorithm for merging two summaries.

\item A theoretical and an empirical analysis of the algorithm.

\item We compare different strategies  (smallest-first, largest-first, greedy-parallel, random) for merging multiple views.

\end{itemize}

Below, we summarize the related works.
Section~\ref{sec:problem-formalization} provides a problem formalization of merging multi-view graph summaries. 
Section~\ref{sec:algorithm} introduces our algorithm to merge two summaries and contains a complexity analysis.
The experimental apparatus is described in Section~\ref{sec:experimentalapparatus}.
An overview of the achieved results is in Section~\ref{sec:results}. 
Section~\ref{sec:discussion} discusses the results before we conclude.

\section{Related Work}
We introduce works on graph summarization and approaches to computing them.
Afterward, we look into multi-view graphs and their applications as an outlook for possible applications for multi-view summaries.
Finally, we discuss the literature on knowledge graph generation because we generate knowledge graphs using GPT-4. 

\textit{Graph summarization.}
Graph summaries can be categorized as either quotient-based~\cite{DBLP:journals/tgdk/ScherpRBCR23}, where vertices are grouped based on shared features, or non-quotient-based, where vertices are grouped according to specific criteria~\cite{DBLP:journals/vldb/CebiricGKKMTZ19}.
In this work, we use the quotient-based definition of structural summaries by~\citet{DBLP:journals/tgdk/ScherpRBCR23}, in which each vertex in the summary corresponds to an equivalence class (EQC) of vertices in the original graph.
We specifically use the quotient-based summary models AC, CC, and ACC introduced by~\citet{DBLP:conf/dexaw/CampinasPCDT12}.
FLUID~\cite{blumericherbyscherp-tcs-2021} is a language and a generic algorithm that flexibly defines and efficiently computes quotient-based structural graph summaries. 
It can express various existing structural graph summaries that are lossless regarding the considered features, such as the vertices’ edge labels, neighbor types, and others. 
FLUID computes graph summaries based on hash functions applied to a canonical order of vertex features.
Another approach for computing graph summaries based on hash functions is SchemEX~\cite{DBLP:journals/ws/KonrathGSS12}, which is a tool for stream-based indexing and schema extraction of linked open data.

We note that, in the literature, there are other terms used for the concept of creating a smaller graph representation while preserving essential information, such as graph condensation~\cite{DBLP:journals/corr/abs-2402-02000} and graph reduction~\cite{DBLP:journals/corr/abs-2402-03358}.  
Other works \cite{shabani2024comprehensive,DBLP:journals/corr/abs-2402-12001} also refer to graph summaries but use a different definition of graph summarization. 

\textit{Multi-view graphs.}
There are already various applications for multi-view graphs in different areas.
An example is representation learning for drug-drug interaction prediction~\cite{DBLP:conf/www/WangMCW21}. 
\citet{DBLP:conf/aaai/LongXC0C022} use multi-view graphs to describe different code views: control flow, data flow, and read-write flow.
These views stem from one connected graph.
\citet{DBLP:conf/aaai/LongXC0C022} use multi-view graph representation for multi-labeled code classification. 
The multi-view source code representation model ACAGNN by~\citet{li2022acagnn} uses the abstract syntax tree, control flow graph, and application programming interface dependency graph to learn multi-view code structure representation and apply it to classify code snippets by functionality.
None of these works addresses summarization of multi-view graphs.

\textit{Knowledge graph generation.}
Knowledge graphs integrate both knowledge and relations into a graph structure, but due to the volume of data, creating them from structured or unstructured data is often difficult.\cite{DBLP:conf/cikm/IglesiasJCCV20}
The SDM-RDFizer~\cite{DBLP:conf/cikm/IglesiasJCCV20} is a tool to generate meaningful knowledge graphs from structured data.
It is fast and has low memory usage, as demonstrated in the Knowledge Graph Creation Challenge~\cite{DBLP:conf/esws/IglesiasV23}, which includes duplicate removal, join execution, and avoiding the generation of empty values.
High-quality knowledge graphs primarily rely on human-curated structured or semi-structured data~\cite{DBLP:conf/emnlp/Ye0CC22}. 
According to~\citet{DBLP:journals/corr/abs-2305-04676}, GPT-3 can already keep up with REBEL~\cite{DBLP:conf/emnlp/CabotN21}, an application for relation extraction from text.
In this work, we use GPT-4~\cite{openai2024gpt4} to generate graphs from the text of news articles.

\section{Problem Formalization}
\label{sec:problem-formalization}

We define the basic notation for multi-view graphs $\mathcal{G}$ and summaries $\mathcal{S}$.
We define graph summary models and graph summarization.
Finally, we describe how to merge two graph summaries.

\subsection{Definition of Graph, Summary Graph, and Multi-views}

We use multi-relational, labeled graphs  $G = (V, E, \ell_V, \ell_E )$ as our overall graph definition. 
These consist of a set of vertices $V$ and a set of labeled edges $E \subseteq V \times V$. 
Each vertex $v \in V$ has a finite set of labels $\ell_V(v) \subseteq \Sigma_V$ and each edge $e = (u,v)$ has a finite set of labels $\ell_E(u,v) \subseteq \Sigma_E$, where $\Sigma_V$ and $\Sigma_E$ are disjoint sets of labels. $ \mathcal{G} = \{ G_1, \ldots, G_n \} $ is a set of graphs, interpreted as providing multiple views of the same underlying graph.
We note that the individual graphs $G_i$ are not necessarily connected (e.\,g.,  view~$2$ in Figure~\ref{fig:flowG}).

A summary graph consists of schema information and payload~\cite{blumericherbyscherp-tcs-2021}. 
The schema is the information used to determine the equivalence of two vertices, i.\,e., to summarize them.
For example, one may use the label set of the outgoing edges of a vertex to find equivalent vertices.
Payloads are computed during summarization and contain additional information about a schema, such as which vertices of the original graph have this schema or where to find it.

We define a summary graph as the combination of schema summary and payload with $S=(V, B, E, \ell_V, \ell_E )$~\cite{blumericherbyscherp-tcs-2021}, where $V$ are the schema summary vertices, $B$ are the payload vertices ($V \cap B = \emptyset$), and $E$ is the set of all edges. 
The summary of a multi-view graph $\mathcal{G}$ is the set $\mathcal{S} = \{ S_1, \dots, S_n \}$, where each $S_i$ is the summary of the corresponding $G_i\in G$.
The label function $\ell_V$ maps each schema summary vertex and payload vertex to a finite set of labels with the restriction that $V$ and $B$ have different labels. 
The function $\ell_E$ labels the edges, and edges from $V$ to $B$ have the label set $\{\text{payload}\}$.

\subsection{Graph Summarization using Equivalence Relations}
\label{sec:summaryModel}

\label{def:summarymodel}
A graph summary model is a pair $M=(\psi, \Lambda)$, where $\psi$ is a function that specifies an equivalence relation on graph vertices and $\Lambda$ is a set of typically task-specific payload functions.
They are computed during summarization, and their output is included in the summary.
The parameter $\psi$ is a function defining the vertex summary of a single vertex $v$ in a graph~$G$, i.e., $\psi(v;G) = S(v)$. 
It induces an equivalence relation on $V(G)$  such that $x \equiv y$ iff $\psi(x; G) = \psi(y; G)$. 
Here, ``$=$'' denotes literal equality of graphs rather than isomorphism.
Thus, we may identify the summary of a vertex with its equivalence class (EQC) in this induced equivalence relation.
$\Lambda$ is a set of functions defining what payload will be attached to each set of vertices $X$ that are equivalent under $\psi$. 
Each function in $\Lambda$ is of the form  $\lambda\colon X \to P$.
Here, $X$ is a subset of vertices of the input graph~$G$ that are equivalent under $\psi$, and $P$ is an appropriate payload domain, such as the natural numbers or set of IRIs.
For example, the count payload is defined by the function $\lambda_\mathrm{count}(X) = |X|$ (see example payload functions in \cite{blumericherbyscherp-tcs-2021}).
We apply the graph summary model to each graph $G \in \mathcal{G}$ to create its summary $S \in \mathcal{S}$.

\label{def:grap_summary}
\label{def:grap_summary_v}
The parameterized function $C_M \colon (v; G) \to S$ computes a representation of the vertex $v \in V$ of the input graph, i.\,e., the EQC of $v$ based on a graph summary model $M$.
The summary~$S$ of a graph~$G$ w.r.t.\@ a summary model $M$ has a vertex for each EQC of each equivalence relation used in the definition of~$\psi$ in~$M$.
In other words, the vertices in $S$ are the EQCs computed by $C_M$ and each EQC is a set of summarized vertices of the input graph.
In consequence, the vertices in $S$ change when the vertices of the input graph change. 
However, in any practical implementation~-- as in ours, see above~-- one distinguishes the schema information and the payload of an EQC~\cite{blumericherbyscherp-tcs-2021}.
Thus, two EQCs from two summary graphs are considered equal iff they represent the same schema information.

For example, the summary model AC~\cite{DBLP:conf/dexaw/CampinasPCDT12}, $\psi$ maps two vertices $u$ and $v$ to the same EQC if they have the same labels on their outgoing edges ($\bigcup_{(u,x)\in E} \ell_E(u,x) = \bigcup_{(v,y)\in E} \ell_E(v,y)$). 
In CC, $\psi$ maps two vertices $u$ and $v$ to the same EQC if they have the same typeset ($ \ell_V(u) = \ell_V(v)$). 
ACC is the combination of AC and CC.
The payload of an EQC in each summary model is the set of mapped vertices. 
We use this payload to identify cases in our algorithm for summary merging.

\subsection{Merging Graph Summaries}
\label{sec:merging summaries}

\begin{figure}[h]
\centering
\includegraphics[width=0.75\linewidth]{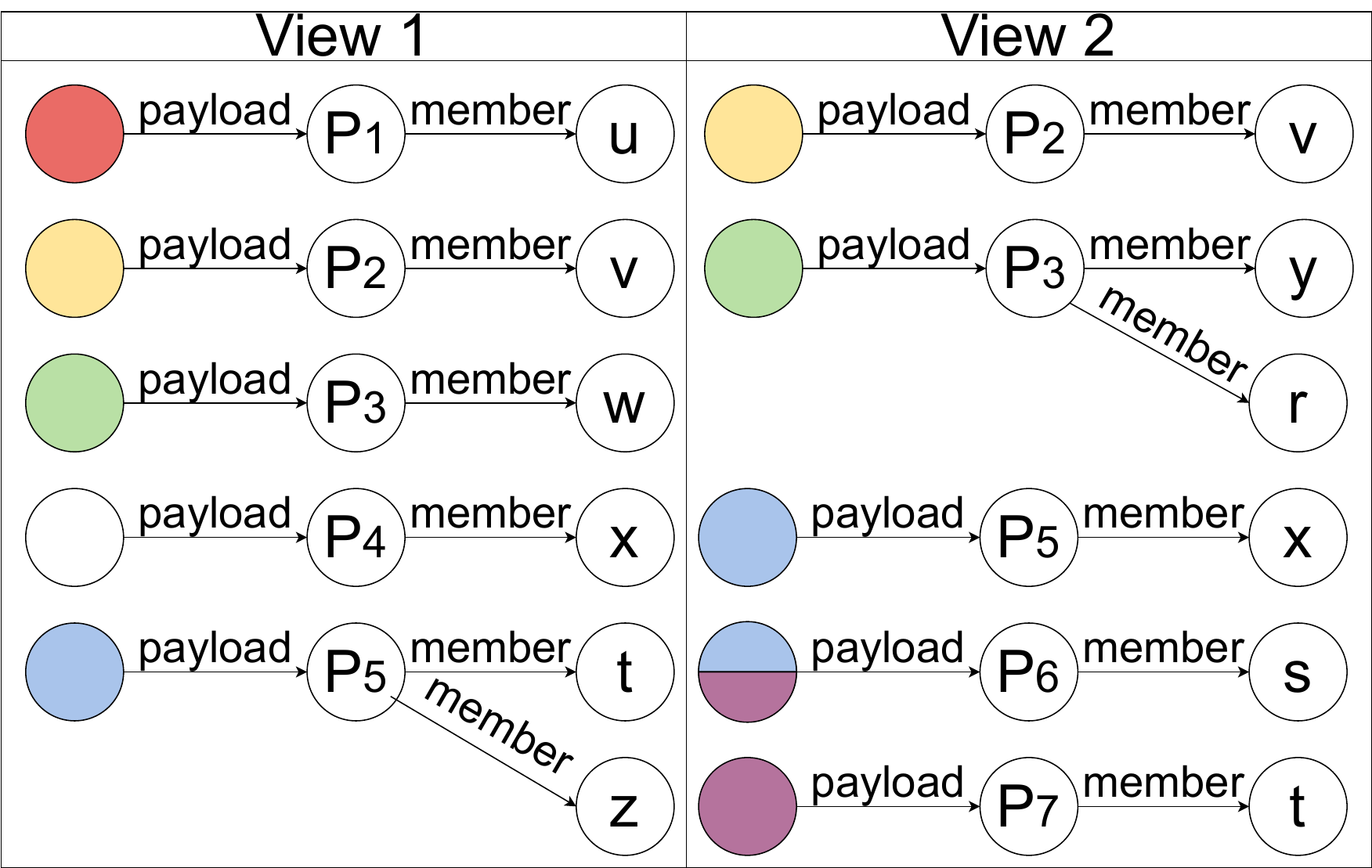}
\caption{The problem of merging two graph summaries (views).
The vertex color represents the EQC under the summary model AC, i.\,e.,  the set of outgoing edges. 
The vertices $t$,  $v$, and  $x$ appear in both views.
}
\label{fig:graphCases}
\end{figure}

We now describe our algorithm for merging summaries $S_1$ and $S_2$ of graphs $G_1$ and~$G_2$.
We merge $S_1$ into~$S_2$, so the algorithm is not symmetric with respect to the two summaries.
We consider three cases for each vertex $v \in G_1$. 
The cases depend on whether $v$ is in one or both graphs and whether the EQC of $v$ is in both summaries.
Figure~\ref{fig:graphCases} shows two summaries that are merged, with vertex colors representing EQCs. 
The payload shows which vertices of the original graph belong to each EQC.
We use this figure to explain the different cases. 

Case~$1$ is trivial.
It applies to vertices that appear only in~$G_1$ and whose EQC appears only in $S_1$ (e.\,g., $u$ in Figure~\ref{fig:graphCases}); and to vertices that appear in both $G_1$ and~$G_2$, with the same EQC in $S_1$ and~$S_2$ (e.\,g., $v$ in the figure).
In this case, the merge does not change the EQC's members, so no payload change is needed and this part of the merge is essentially just a union.

Case~$2$, the payload merge case, applies to vertices that are only in $G_1$, but whose EQC exists in both summaries.
In Figure~\ref{fig:graphCases}, $w$ is in this case: it is only in~$G_1$, but the green EQC appears in both summaries.
In this case, we may need to adapt the payload of the EQC in the merged summary to contain information about all its members from both $G_1$ and~$G_2$.
Whether an adaptation is necessary depends on the payload functions.
In the figure, the payload is just a $\text{member}$ edge to each vertex in the EQC, so no adaptation is needed beyond taking the union of the vertex summaries as in case~$1$.
However, if the payload was the number of vertices in the EQC, this would need to be updated from $1$ in $S_1$ and $2$ in $S_2$ to $3$ in the merged summary.

Case~$3$, the EQC merge case, is for vertices that appear in both graphs but have a different EQC in each, e.\,g.,  $x$ and~$t$ in Figure~\ref{fig:graphCases}. 
In this case, the two EQCs of $x$ must be combined.
If this combined EQC already exists, its payload is adapted; otherwise, the payload is created along with the new EQC.

Independent of the cases, a fundamental assumption of merging two summaries  $ S_1 $ and $ S_2 $  is that their EQCs have consistent identifiers. 
That means if  $S_1 $ and $ S_2 $ contain vertices representing the same schema information, those vertices have the same identifier.
In theory, this can be achieved by the identifier of an EQC being the schema it represents. 
A possible implementation of the identifier for calculating the EQCs would be a hash of the schema structure, assuming no collisions, as with SchemEX~\cite{DBLP:journals/ws/KonrathGSS12}.

\section{Algorithm for Pairwise Merging}
In this section, we introduce an algorithm for merging two summaries and explain it step by step.
We analyze its runtime, and we show that it generalizes to merging $n$ summaries.
\label{sec:algorithm}
\begin{figure}
\centering
\includegraphics[width=0.47\textwidth]{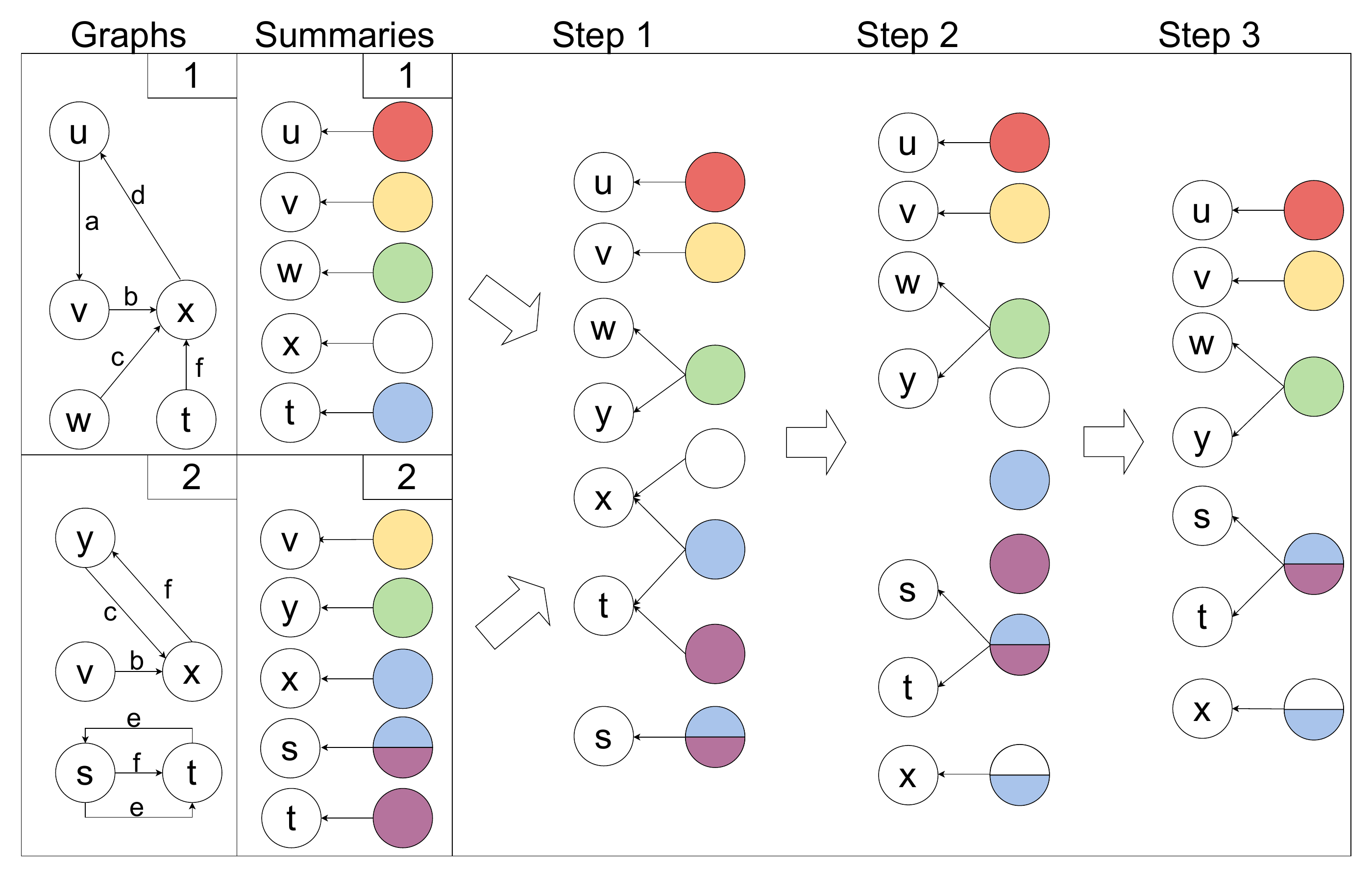}
\caption{Merging two summaries. 
The color of the vertices indicates the EQC. Payload vertices are omitted for clarity. Here, the payload is directly attached to the EQC. 
}
\label{fig:flowG}
\end{figure}

\begin{algorithm}
  \caption{An algorithm for merging two summaries $S_1$ and $S_2$ based on the summary model $M=(\psi, \Lambda)$.
    \label{alg:merging}}
  \begin{algorithmic}[1]
  \Statex \textbf{Input:} two summaries  $S_1=(V_1, B_1, E_1)$ and $S_2=(V_2, B_2, E_2)$ and a definition of the summary model $M=(\psi, \Lambda)$.
    \Function{MergeSummaries}{$S_1, S_2,\psi,\Lambda$}
    \Statex \hspace{0.5cm}// \textit{Step 1} 
    \Let{$S$}{$S_1 \cup S_2$} \Comment{Case~$1$}   
    \Statex \hspace{0.5cm}// \textit{Step 2} 
    \For{$m \in \:$\Call{GetMembers}{$S_1$}}
         \If{$m \in \:$\Call{GetMembers}{$S_2$}}
                \Let{$c_1$}{\Call{GetEQC}{$S_1,m$}}
                \Let{$c_2$}{\Call{GetEQC}{$S_2,m$}}
                \If{$c_1\neq c_2$}   \Comment{Case~$3$}               
                    \Let{$S$}{\Call{CombineEQCs}{$S,c_1,c_2,m,(\psi, \Lambda)$}}
                \EndIf
        \EndIf
        
    \EndFor 
    \Statex \hspace{0.5cm}//  \textit{Step 3} 
    \For{$v \in $ \Call{GetAllEQCs}{$S$}} 
        \If{\Call{HasMembers}{$S,v$}}
        \Let{$S$}{\Call{AdaptPayload}{$S,v,\Lambda$} }\Comment{Case~$2$} 
        \Else
        \Let{$S$}{\Call{RemoveEmptyEQC}{$S,v$}}  \Comment{Case~$3$} 
        \EndIf
        
    \EndFor
   
      \State \Return{$S$}  
    \EndFunction
  \end{algorithmic}
\end{algorithm}

\begin{algorithm}
\caption{Auxiliary functions used by Algorithm~\ref{alg:merging}. 
}
    \label{alg:Aux}
  \begin{algorithmic}[1] 
  \Statex //  \textit{Returns all the member vertices in a summary $S$, i.\,e.,  all vertices in $G$.} 
    \Function{GetMembers}{$S$}
    \State \Return{$\{m \mid (*,\text{member},m) \in S.E\}$}
    \EndFunction
    \Statex
    \Statex //  \textit{Returns the unique EQC of vertex $m$ in summary $S$.} 
    \Function{GetEQC}{$S,m$}
    \State \Return{$c\text{ s.t. } \exists p\,(c, \text{payload}, p), (p,\text{member},m) \in S.E$}
    \Statex
    \EndFunction  
    \textit{// Merges the two EQCs $c_1$ and $c_2$  using the summary model $M=(\psi, \Lambda)$.}
    \Function{CombineEQCs}{$S,c_1,c_2,m,M$}
         \Let{$S.E$}{$S.E\backslash \{(s,\text{member},m)\mid (s,\text{member},m) \in S.E\}$}
            \Let{$t_E$}{$\{(m,p,o)\mid (c_1,p,o)  \in S.E \lor (c_2,p,o) \in S.E \}$}
            \Let{$t_V$}{$\{v \mid (*,*,v)  \lor  (v,*,*) \in t_E\}$}
            \Let{$G$}{$(t_V,t_E)$}
            \Let{$T$}{$C_M(m;G)$} \Comment{see Section~\ref{sec:summaryModel}} 
\State \Return{$S \cup T$}  
    \EndFunction
    \Statex
  \Statex //  \textit{Returns all EQCs of summary $S$.} 
    \Function{GetAllEQCs}{$S$}
    \State \Return{$\{c \mid (c, \text{payload}, p) \in S.E\}$}
    \Statex
    \EndFunction

\Statex //  \textit{Returns whether an EQC $c$ in the summary $S$ has members.} 
    \Function{HasMembers}{$S,c$}
    \Let{$t$}{|$\{m \mid (c, \text{payload}, p) ,(p,\text{member},m) \in S.E\}|$}
    \State \Return{$t > 0$}
   \Statex
    \EndFunction

\Statex //  \textit{Removes EQC $c$ from summary $S$.} 
    \Function{RemoveEmptyEQC}{$S,c$}
    \Let{$S.V$}{$S.V \backslash \{c\}$}
    \Let{$S.B$}{$S.B \backslash \{b \mid (c,\text{payload},b) \in E\}$}
    \Let{$S.E$}{$S.E \backslash \{(c,p,o) \mid (c,p,o) \in E\}$}
\State \Return{$S$}  
\EndFunction
  \end{algorithmic}
   
\end{algorithm}

\subsection{Algorithm}

Let $S_1$ and $S_2$ be summaries of graphs $G_1$ and $G_2$, respectively, which may intersect.
Algorithm~\ref{alg:merging} takes these summaries as input in the form of $S_i=(V_i, B_i, E_i)$ ($i\in \{1,2\}$) to merge them.
Specifically, the algorithm merges summary $S_1$ into~$S_2$, so it is not symmetric w.r.t.\@ the two summaries.
For clarity, the label functions $\ell_V$ and $\ell_E$ are omitted in the algorithms.
The edges, payloads, and vertices of a summary $S$ are accessed in our functions as $S.E$,  $S.B$, and  $S.V$, respectively.
We use ``$*$'' as a wildcard character.

Algorithm~\ref{alg:merging} has three steps, which are also depicted in Figure~\ref{fig:flowG}.
In step $1$, the two summaries are simply merged by taking $S=(V_1 \cup V_2, B_1 \cup B_2, E_1 \cup E_2)$ as if all vertices belong to case~$1$, the trivial EQC merge case (a vertex either is in both graphs and has the same EQC in both, or it and its EQC each appear in just one graph).
The next two steps correct errors caused by case~$2$, the payload merge case (a vertex appears only in one graph, but the other graph contains vertices of the same EQC), and case~$3$, the EQC merge case (a vertex appears in both graphs but in different EQCs).

In step $2$, case~$3$ is detected, and EQCs are merged if necessary. 
In this step, the auxiliary functions in Algorithm~\ref{alg:Aux} are used.
\textproc{GetMembers}($S$) returns all members of the EQCs in a summary $S$. 
If two summaries $S_1$ and $S_2$ have the same member $m$ but \textproc{GetEQC}($S_1,m$) and \textproc{GetEQC}($S_2,m$) return different EQCs, then the schemas of the two EQCs are merged to create the EQC of  $m$ for the merged summary.
Two EQCs $c_1$ and $c_2$ are combined by \textproc{CombineEQCs}($S,c_1,c_2,m,\psi$). 
It removes $m$ as a member of $c_1$ and $c_2$ and adds it as a member to the newly calculated EQC.

In step $3$, the payloads are corrected, and EQCs without members are removed. 
This step also uses auxiliary functions in Algorithm~\ref{alg:Aux}.
All EQCs in $S$ are returned by \textproc{GetAllEQCs}($S$). 
\textproc{HasMembers}($S,c$) determines for each EQC $c$ if it has any members in $S$.  
If an EQC $c$ has at least one member, the payload may need to be adapted due to case~$2$.  
This is done by \textproc{AdaptPayload}($S,c,\Lambda$). 
The function \textproc{AdaptPayload} depends on the summary model's set of payload functions $\Lambda$ (see Section~\ref{sec:merging summaries}). In our case, it just returns~$S$: no adjustments are needed for our payloads because they only contain the members of the EQCs.  
A vertex either stays a member of an EQC or changes the EQCs due to case~$3$.
If an EQC $c$ has no members, it is removed by \textproc{RemoveEmptyEQC}($S,c$). 
This case only happens when, due to case~$3$, all members are removed from an EQC. 
After step $3$, Algorithm~\ref{alg:merging} returns the merged summary.

\subsection{Complexity Analysis}
\label{sec:complexity}
Algorithm~\ref{alg:merging} merges two summaries, assuming that members of each EQC are already known.
The worst-case running time of  Algorithm~\ref{alg:merging} mainly depends on the size of $E_1$ and $E_2$. $E$~is the combination of $E_1$ and $E_2$. 
We assume that in the worst case, $E$ is bigger than $V$ and $B$, and $E_1$ and $E_2$ are about the same size $\ell$. 
This gives, $\ell\leq|E|\leq 2\ell$.

In step $1$, the vertices, payload, and edges of the summaries are merged. 
This takes time $\mathcal{O}(|E|)$, assuming we use a hash table and can assume constant key size, independent of $|E|$.

Step $2$ tests for case~$3$.
All the members in $S_1$ are searched, and then the algorithm looks in $S_2$ to see if a member of $S_1$ also appears in $S_2$. 
If a member $m$ appears in both summaries, then the algorithm searches for its EQCs in $S_1$ and $S_2$. 
If $m$ has different EQCs, then they are merged, and $m$ is added as a member to the merged EQC and removed as a member from its original EQCs. 
The merging of EQCs $c_1$ and~$c_2$ can be done in time $\mathcal{O}(|E|)$ by a linear scan of $E$, $V$, and $B$, replacing the elements of $c_1$ and $c_2$. 
Again, we assume that $E$ is the largest component of $S$. 
Step $2$ takes time $\mathcal{O}(|E|^2)$, as it takes time $\mathcal{O}(|E|)$ to find all members that are in two EQCs and $\mathcal{O}(|E|)$ to merge two EQCs.

In step $3$, each EQC's payload is searched and corrected. 
This has running time $O(|V| \, |E|)$ because the number of EQCs is at most~$|V|$, and for each EQC, $E$ is searched and altered. 
Overall, the worst-case running time of the algorithm is  $\mathcal{O}(|E|^2)$.

The running time reduces to $\mathcal{O}(|E|)$ under the assumption that all information needed from $E$ can be saved in advance so that it can be accessed in time $\mathcal{O}(1)$, and that we always have hash tables with constant-sized keys. 
Therefore, we can merge sets in linear time and alter sets in constant time. 
This assumption reduces the running time of all three steps to $\mathcal{O}(|E|)$.
This is because EQCs can be merged in constant time since elements can be added and removed in constant time, and all information needed can be accessed in constant time.
Therefore, the algorithm has worst-case running time $\mathcal{O}(|E|)$. 

\subsection{Merging $n$ Summaries}
\label{sec:n-summaries}
With repeated application, Algorithm~\ref{alg:merging}  can merge $n$ summaries into one summary. 
One key element for the runtime of merging $n$ summaries is choosing which pairs are merged first. 
In our complexity analysis, we already showed that the runtime of the algorithm depends on the number of edges. 
Hence, it makes sense to always merge the two smallest summaries by the number of edges, which we call the smallest-first strategy. 
Nevertheless, that does not necessarily mean that choosing the two largest summaries by the number of edges for merging, namely the largest-first strategy, results in a worse runtime. 
Another possible strategy is to choose the pairs randomly. 

An additional benefit of our proposed algorithm is that it can merge different pairs simultaneously. 
So, a strategy using this fact may reduce the runtime of merging $n$ summaries. 
Under the assumption that we have as many CPUs as initial summary pairs, we can merge all pairs simultaneously, starting with pairs of similar sizes.
When a pair of summaries has been merged, we must wait for the next pair to be merged before the two new summaries can themselves be merged.
Our strategy is to merge as soon as there are two summaries available to merge, rather than waiting in the hope of getting a better combination of summaries; therefore, we call this strategy the greedy-parallel strategy.
The time for merging all summaries can be explained recursively. 
The time it takes to create a summary $i$ is the time it takes to merge the pair $(i-1,i-2)$, plus the longer of the times taken to produce the summaries in the pair. 
The initial summaries are part of the input, so take no time to produce.

\section{Experimental Apparatus}
\label{sec:experimentalapparatus}

\subsection{Datasets}
\label{sec:datasets}

\begin{table*}
    \centering
     \caption{Characteristics of the datasets. The graphs in a dataset provide either a different view of the same topic (one domain) or are about different topics (many domains). 
    They also have either the same source (single source) or different sources (multiple sources).
    Source objects (``Src.\@ Obj.'') refers to the number of multi-view graphs in a dataset.}
    \label{tab:data}
    \begin{tabular}{|l|rrr|rrr|rrr|} \hline
         Dataset & Total \# Edges  & Total \# Vertices  &  Src.\@ Obj. & Views & Sources &  Domains & EQCs in AC & EQCs in CC & EQCs in ACC\\ \hline
         BTC 2019 & $2,155,856,033$ & $42,059,543$ & $1$  & $101$  & Multiple  & Many & $400,297$& $560,984$& $1,047,644$\\ 
         \code & $3,550,689,526$  & $404,734,527
$  & $329,797$  &$3$& Single &  One & $3,039$ & $81$ & $3,416$\\ 
        \news & $30,156$  &  $10,892$& $15$  & $3$& Multiple  & One  & $4,054$ & $1,284$ & $4,946$\\\hline
    \end{tabular}
   
\end{table*}

We use three different datasets to test our algorithm.
For each, we calculate the AC, CC, and ACC summaries. 
We represent graphs in the RDF model. 
RDF is a W3C standard~\cite{w3c-rdf} that models relations in the form of subject--predicate--object triples $(s,p,o)$. 
We use line-based formats for encoding the triples as the syntax to store our graphs. 
Table~\ref{tab:data} shows the datasets' properties.

\paragraph{Billion Triple Challenge 2019 Dataset}
The BTC 2019 dataset~\cite{DBLP:conf/semweb/HerreraHK19} contains $2,155,856,033$ edges collected from $2,641,253$ RDF documents on 394 pay-level domains.  
However, we consider 101 different domains and, therefore, 101 views from multiple sources and with many domains. 
The domain ``other'' aggregates the $294$ pay-level domains that are not in the top 100 by number of triples. 
Considering ``other'' as one view results in only a small error because only $97,643$  of the $2,155,856,033$ edges are in this view~-- less than $0.005\%$.
The whole dataset consists of $256,059,356$ triples,  containing $38,156$ predicates and $120,037$ classes.

\paragraph{CRAN Social Science Three Views Dataset}
We created the \code dataset using the flowR framework\footnote{\url{https://github.com/Code-Inspect/flowr}} by~\citet{flowR}. 
This framework creates three views: a normalized abstract syntax tree representation, a control flow, and a data flow of an R~program.
flowR reads the abstract syntax tree returned by R's parser. 
The control flow representation of flowR does not show what is executed next; it shows what has to be executed before. 
This seems counterintuitive at first, but it still models the dependencies between the elements.
All graphs produced by flowR are returned in JSON format.
We use our own program to convert them into RDF graphs.

We apply flowR to the $347,867$ R-files of the CRAN packages and social science files provided by Sihler and Tichy~\cite{flowR}.
In contrast to the BTC 2019 dataset, each multi-view graph here has a single source, namely the R-file. Each file has three different views from one domain, the source code. 
In total, $329,797$ files were successfully parsed, $956$ files could not be parsed, and for $17,121$ files, not all views could be created. 

\paragraph{International News Coverage 2023 Dataset}
To further diversify our datasets, we created the \news dataset.
For this dataset, we used GPT-4~\cite{openai2024gpt4} to generate RDF graphs of news articles from the web.  
We decided to focus on five different columns, namely \textit{Sports}, \textit{Entertainment}, \textit{Social Media}, \textit{Science}, and \textit{Health}.
For each column, we have three topics. 
The exact topics for each column are found in Table~\ref{tab:coloums}.
For each topic, we searched for ten articles on three international news outlets, namely \textit{aljazeera.com}, \textit{euronews.com}, and \textit{cnn.com}.
We chose these three news domains to cover a wide range of geopolitical perspectives. 
For the article search, we used the website's search engine and sorted the results by relevance. 
We selected the first ten, excluding videos. 
These ten articles show how a broadcaster handles a topic.
We downloaded the articles from the outlets using the newspaper framework.\footnote{\url{https://github.com/codelucas/newspaper}}
We used GPT-4 to create an RDF graph in turtle syntax for each article. 
Specifically, we used \textit{gpt-4-1106-preview} on 15--16 November 2023 to create the graphs, with the prompt,
``Create an RDF graph in Turtle syntax that contains the information in the following text. 
Use no long strings as abbreviations''.
The temperature is set to $0$. 

Since the graph generation is imperfect, we post-processed to remove syntax errors in the 450 graphs.
The most common cause is invalid characters in RDF triples, which causes syntax errors in 73 files. In 39 files, prefixes are missing, and in 16 files, triples do not have three values. Problems with an RDF list are found in ten files. Comments (nine files) and wrong typing of literals (four files) also cause syntax errors.
These errors are only minor and were removed by hand.
Afterward, the RDF graphs for the same topic and broadcaster were combined to create one RDF graph. 
Hence, we have 45 graphs covering 15 topics from three broadcasters.
Each multi-view graph has multiple sources (different broadcasters) and is about the same topic (one domain).

\begin{table}
    \centering
    \caption{Five news columns and with each three topics.}
    \label{tab:coloums}
    \begin{tabular}{|l|l|}\hline
    Column & Topic \\ \hline
    Sports & FIFA, NBL, NFL \\\hline
   Entertainment &  Disney, Nintendo, Warner Bros. \\\hline
   Social Media & Facebook, Twitter,  TikTok   \\\hline
   Science &  CERN, Nobel Prize, Stanford University \\ \hline
   Health  &  World Health Organization, Cigarettes, Obesity \\ \hline

         \hline
    \end{tabular}
    
\end{table}

\subsection{Procedure}
\label{sec:procedure}
We perform our experiment once for the summary models AC, CC, and ACC. 
We use a hash function to calculate each graph's summary. 
We save the EQCs with their payloads as triples.
Then, we apply our Algorithm~\ref{alg:merging} for pairwise merging the multi-view summaries. 
We exclude pairs consisting of the same summary, and for two summaries $S_1$ and $S_2$, we consider both merge pairs $(S_1, S_2)$ and $(S_2, S_1)$. 
We consider both combinations because the case distribution is not symmetric.
Both the number of members and the number of cases $1$ and~$2$ depend on the first summary. 
The number of cases~$3$ is the same for both combinations because, for case~$3$, the vertex has to be in both summaries.
In the BTC 2019 dataset, we apply it to the $10,100$ ordered pairs of distinct summaries that arise from the 101 views.
For the code graphs, we only apply Algorithm~\ref{alg:merging} to pairs of graph views of the same code graph.
We have three views, resulting in six pairs per R-file: 
$1,978,782$ pairs in total. 
For the news graphs, we merge $90$ pairs.
We track how long each merge takes.

Additionally, we merge all summaries of BTC 2019 into one summary using our algorithm.
Because the order of pairwise merging influences the running time, we merge the summaries of the BTC 2019 dataset using four different strategies, namely smallest-first, largest-first, and greedy-parallel, and random (see Section~\ref{sec:n-summaries}).
For the first three strategies, we run our algorithm six times.  
The first run is ignored (used as a warm-up) and the running times of the other five runs are averaged.
We run the random strategy ten times and average the results.
All experiments are conducted on an 8-CPU AMD EPYC 7F32 with 64 cores and 1.96 TB of RAM.

We measure the time that pairwise merging takes in our implementation,  
and calculate the Pearson correlation between the pairwise merge times and the three functions $|E|$, $|E|\log |E|$, and $|E|^2$. 
We also perform linear regression and calculate the coefficient of determination~$R^2$, and calculate the average time and standard deviation for the different strategies for merging the BTC 2019 summaries.

\section{Results}
\label{sec:results}

\begin{table}
\caption{ Pearson correlation between merge time and function of $E$ for each dataset. The highest value for a correlation between a summary and function of $E$ per dataset is in bold. All \textit{p}-values are lower than 0.05.}
    \label{tab:Correlation}
    \begin{tabular}{|l|l|r|r|r|} \hline
       \vtop{\hbox{\strut Summary}\hbox{\strut model}}   & \vtop{\hbox{\strut Function}\hbox{\strut of $E$}}  & BTC 2019    & CSS3V & INC 2023\\ \hline
           & $|E|$            & \textbf{0.931} & \textbf{0.774} & 0.429 \\
        AC & $|E|\log |E|$  & \textbf{0.931} & 0.773 & 0.429 \\
           & $|E|^2$          & 0.875 & 0.538 & \textbf{0.431} \\ \hline
           &  $|E|$           & 0.787 & 0.716 & \textbf{0.921} \\
        CC & $|E|\log |E|$  & \textbf{0.788} & \textbf{0.719} & \textbf{0.921}\\
           & $|E|^2$          & 0.751 & 0.504 & 0.914\\ \hline
           &  $|E|$           & 0.957 & \textbf{0.866} & \textbf{0.459}\\
        ACC& $|E|\log |E|$  & \textbf{0.959} & 0.865 & \textbf{0.459}\\
           & $|E|^2$          & 0.946 & 0.585 & 0.454\\ \hline
    \end{tabular}
    
\end{table}

\subsection{Billion Triple Challenge 2019 Dataset}

The correlation between the merge time of the pairs and the functions is shown in the left column of Table~\ref{tab:Correlation}. 
For CC and ACC, the Pearson correlation is highest for $|E|\log |E|$.
For AC, the Pearson correlations for $|E|\log |E|$ and $|E|$ are both $0.931$. 
The Pearson correlation for $|E|^2$ is always lower than for the other two. 
For all functions, the Pearson correlation is lower for CC than for the other summary models.
We also computed a linear regression between the functions and the times.
The results (plots in in the Appendix, Figure~\ref{fig:Regreesion}) show that the regression models of $|E|$ and $|E|\log |E|$ are similar, and their coefficients of determination are the highest. 
The coefficients of determination are for CC lower than for the AC and ACC.

The average runtimes for the different strategies for merging the BTC 2019 dataset are in Table~\ref{tab:MergeAll}. 
For all summary models, merging the smallest two summaries (by number of edges) is the fastest. 
For AC and CC, the parallelization also has the best time but with a higher standard deviation. 
The largest-first strategy results in the worst time for all summary models. 
Even randomly merging summaries is faster.
Runtimes were measured with exclusive access to the machine and by conducting a warm-up run.
Thus, observed differences are due to the influence of the operating system.

\subsection{CRAN Social Science Three Views Dataset}

The Pearson correlation between the merge time of the pairs and the functions is shown in the middle column of Table~\ref{tab:Correlation}. 
For AC and ACC, the Pearson correlation between the merge time and $|E|$ is the highest. 
For CC, the correlation is higher for $|E|\log |E|$. 
For all summaries,  the Pearson correlation for $|E|^2$  is over $20\%$ lower than for the other functions. 
This is the highest difference of all datasets. 

The linear regression models for $|E|$ and $|E|\log|E|$ have the same coefficient of determination $R^2$ for AC and CC; for ACC, $R^2$ is higher for $|E|$ than for $|E|\log|E|$.
The models for $|E|^2$ do not fit the data as well as the other models and have lower coefficients of determination for $|E|^2$.

\subsection{International News Coverage 2023 Dataset}

In Table~\ref{tab:Correlation}, the values of Pearson correlation in the right column are similar within each summary model. 
For AC, the Pearson correlation for $|E|^2$ is slightly higher than for the other functions. 
For CC and ACC, $|E|$ and $|E|\log |E|$ have the highest correlation. 
The values of the Pearson correlation for CC are over $45\%$ higher than for AC and ACC.

For each summary model, the three linear regression models have similar coefficients of determination.
The coefficients for $|E|$ and $|E|\log|E|$ are the same in each case, and the coefficient for $|E|^2$ differs by less than $0.01$.
The coefficients of determination are highest for CC and more than $0.45$ lower for AC and ACC.

\begin{table}
    \centering
    \caption{Runtime (in minutes) of merging all $101$ summaries of the BTC 2019 dataset ($10,100$ merge pairs) with different strategies. The results are averaged over ten runs for the random strategy and over 5 runs for the other strategies.}
    \label{tab:MergeAll}
    \begin{tabular}{|l|rrr|}\hline
        Strategy &  Time for AC &  Time for CC   &  Time for ACC  \\ \hline
        Smallest-First & $12.0$  $\pm$ $0.48$  & $9.92$  $\pm$ $0.59$ & $15.3$ $\pm$ $0.79$ \\
        Largest-First & $406.5$ $\pm$ $18.2$ & $426.2$ $\pm$ $34.5$ & $521.5$  $\pm$ $42.9$ \\
        Greedy-Parallel & $12.0$ $\pm$ $1.30$ & $10.7$ $\pm$ $0.52$ & $15.3$  $\pm$ $1.22$\\
        Random & $55.0$ $\pm$ $10.0$  & $45.9$ $\pm$ $12.3$ & $74.8$  $\pm$ $15.3$\\ \hline
    \end{tabular}
    
\end{table}

\section{Discussion}
\label{sec:discussion}
To the best of our knowledge, we are the first to introduce an algorithm to merge two summaries. 
Our Algorithm~\ref{alg:merging} considers the three possible cases described in Section~\ref{sec:merging summaries}.
In the following, we discuss the key insights from our algorithm and experiments.

\subsection{Runtime Complexity and Measurements}
The theoretical running time of our algorithm is $\mathcal{O}(|E|^2)$. 
However, if we adhere to certain assumptions, such as saving all required information from $E$ in advance and using hash tables with keys of constant size, the running time is reduced to $\mathcal{O}(|E|)$.
Algorithm~\ref{alg:merging}'s runtime is an upper bound for the time complexity of the problem of merging two summaries.
Thus, the time complexity of merging is $\mathcal{O}(|E|^2)$.
We also have a lower bound of $\Omega(|E|)$ since it is not possible to merge two summaries without reading them both fully.

Table~\ref{tab:Correlation} shows that the Pearson correlation between $|E|$ and the runtime of merging is either the best of all functions or is only less than $0.03$ behind the best.
Sometimes, the Pearson correlation for $|E|\log |E|$ is the highest. 
In general, $|E|^2$ has the weakest correlation.
However, except for the AC and ACC of the news dataset, there is a strong correlation between the time and the functions of $|E|$.
The \news dataset's poor results can be attributed to the limited number of $90$ pairs, as even a small number of outliers can have a significant impact. 
Overall, a strong correlation exists between the merge time and $|E|$.
The linear regressions further support this, with the coefficients of determination $R^2$ for $|E|$ always being the highest and the lowest $R^2$ being achieved for $|E|^2$. The lines of $|E|\log |E|$ and $|E|$ are similar for the BTC 2019 and \news datasets, with only a noticeable difference between the lines for the \code dataset.
Details can be found in the appendix.

The test results support our theoretical analysis that the runtime of $\mathcal{O}(|E|^2)$ can be improved.
Our empirical evaluation indicates that $|E|$ and $|E|\log |E|$ are better predictors of the measured merged time than $|E|^2$.
However, neither the regression nor the correlation provides conclusive evidence as to whether the runtime is $\mathcal{O}(|E|)$ or $\mathcal{O}(|E|\log |E|)$.

\subsection{Strategies for Merging $n$ Summaries}
Table~\ref{tab:MergeAll} indicates that the optimal strategy for merging multiple summaries is to combine the two summaries with the least number of edges.
Conversely, merging those with the greatest number of edges results in the longest processing time, which is, on average, over 30 times longer than the smallest-first strategy. 
Even the greedy-parallel strategy is worse than the smallest-first strategy. 
This is due to the varying sizes of summaries in the BTC 2019 dataset. For instance, in the BTC dataset, the two largest summaries are larger than the rest of the summaries combined, resulting in longer merge times.
Overall, the smallest-first strategy is a good strategy. 
Nevertheless, if the summaries are similar in size, the greedy-parallel strategy can be faster. 

\subsection{Generalizability}
Our algorithm is designed to merge structural summaries as long as the members of an EQC are known.
Only the helper functions need to be updated to apply the algorithm to other summary models or payload functions. 
A generic solution for merging summaries defined in a language like FLUID~\cite{blumericherbyscherp-tcs-2021} would also be possible.
Our experiments are conducted on three datasets and three summary models to demonstrate the applicability of our algorithm.
As demonstrated in Section~\ref{sec:results}, our algorithm is also capable of merging $n$ summaries.

\subsection{Limitations and Future Work}
We only consider summary models with outgoing edges.
Future work may use summary models considering incoming edges, which is straightforward to add.
A limitation of our algorithm is that it primarily focuses on structural graph summaries, particularly those based on quotients.
Future work may explore extending the algorithm to handle non-structural features. 
Furthermore, we only consider simple merging strategies like the smallest-first strategy. 
In this future work, a deeper exploration into strategies that balance time and space complexity or that account for specific graph topologies may be possible. 

\section{Conclusion}
\label{sec:conclusion}
We formalize the problem of merging summaries and provide two new datasets with different multi-view graphs.
In total, we have three multi-view graph datasets with different characteristics.
We introduce a summary model independent algorithm for merging two summaries.
Our theoretical analysis indicates that the worst-case complexity of the problem is $\mathcal{O}(|E|^2)$ without any assumptions.
However, with certain assumptions, such as having all necessary information saved in advance and having hash tables with constant-sized keys, the complexity is reduced to $\mathcal{O}(|E|)$.
Our empirical analysis supports the argument that the problem's complexity can be reduced. 
We also investigate the process of merging $n$~summaries. Our experiments demonstrate that the smallest-first strategy is the best choice.

\section*{Acknowledgements}
This paper is the result of the first author's Master's thesis.
This research is co-funded by the CodeInspector project (No. 504226141) 
of the DFG, German Research Foundation.

\bibliographystyle{IEEEtranN}
{\footnotesize
\bibliography{main}}
\clearpage
\onecolumn 
\appendix\label{sec:appendix}
\begin{figure}[H]
     \centering
     \begin{subfigure}[b]{0.3\textwidth}
         \centering
         \includegraphics[width=\textwidth]{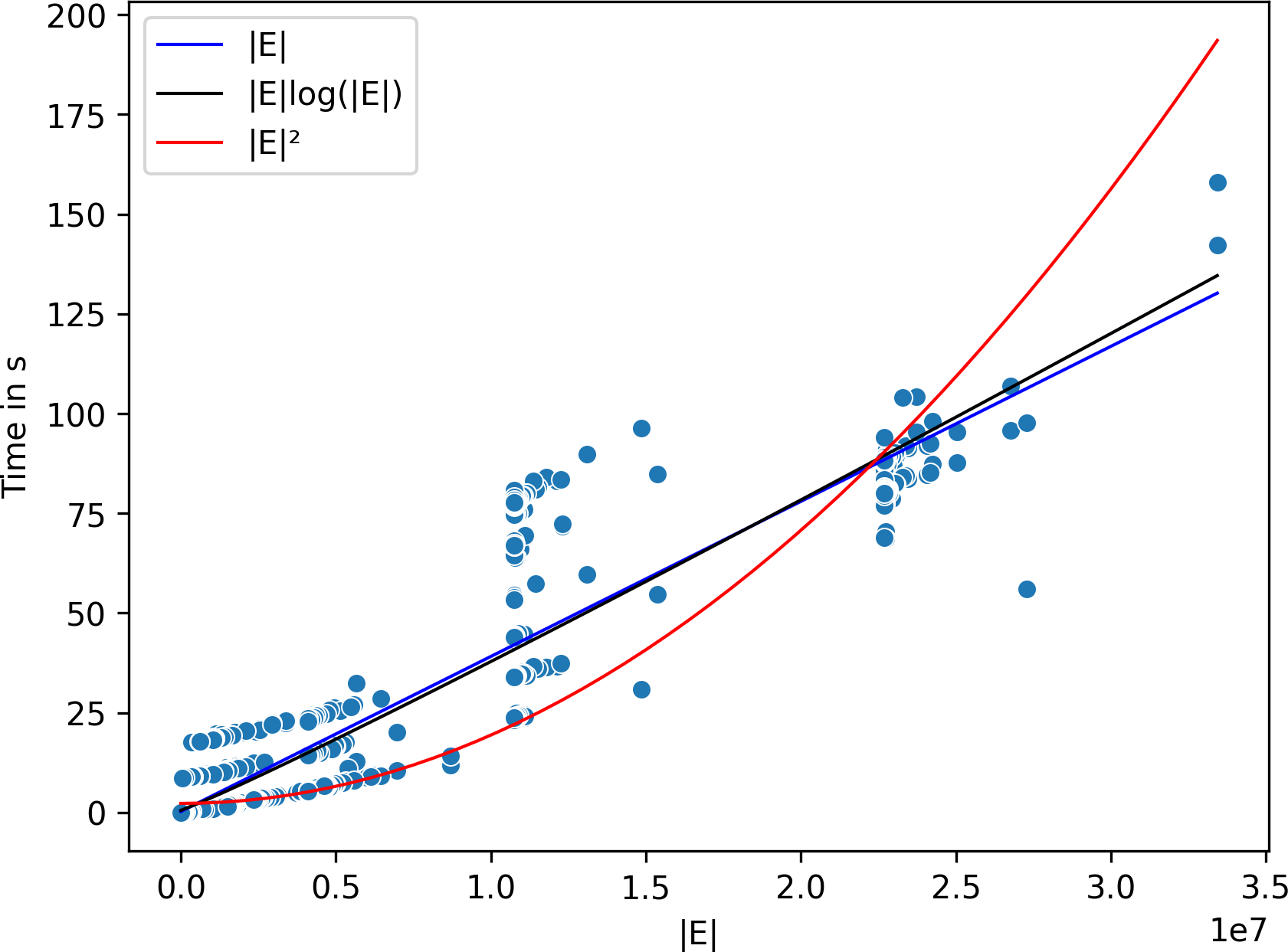}
         \caption{Model: AC, Dataset: BTC 2019 \\
         \textcolor{blue}{$R^2 = 0.93$}, 
         \textcolor{black}{$R^2 = 0.93$}, 
         \textcolor{red}{$R^2 = 0.88$}}
         \label{fig:RegBTC-AC}
     \end{subfigure}  \hfill
     \begin{subfigure}[b]{0.3\textwidth}
         \centering
         \includegraphics[width=\textwidth]{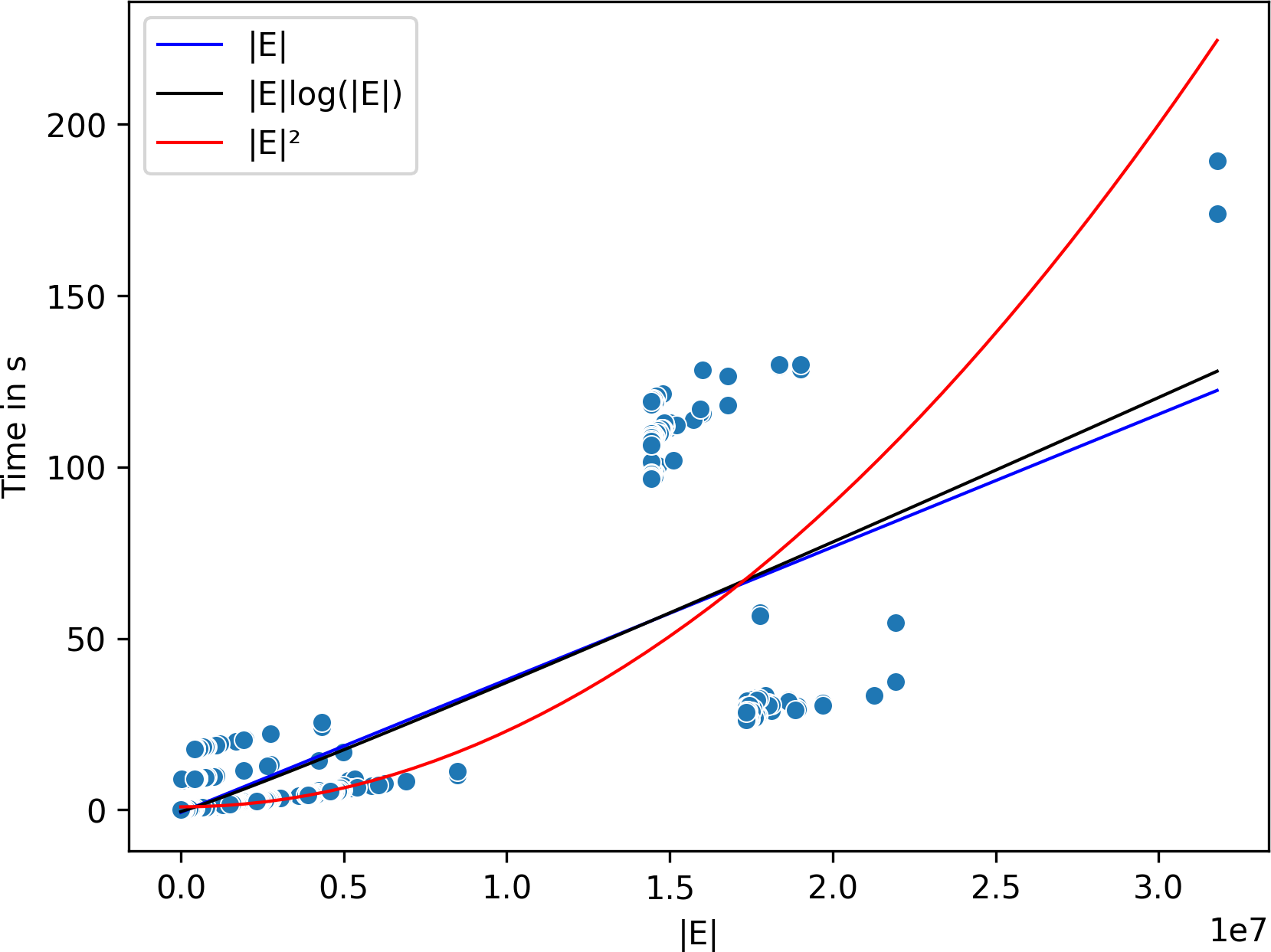}
         \caption{Model: CC, Dataset: BTC 2019\\
         \textcolor{blue}{$R^2 = 0.79$}, 
         \textcolor{black}{$R^2 = 0.79$}, 
         \textcolor{red}{$R^2 = 0.75$ }}\label{fig:RegBTC-CC}
     \end{subfigure}  \hfill
     \begin{subfigure}[b]{0.3\textwidth}
         \centering
         \includegraphics[width=\textwidth]{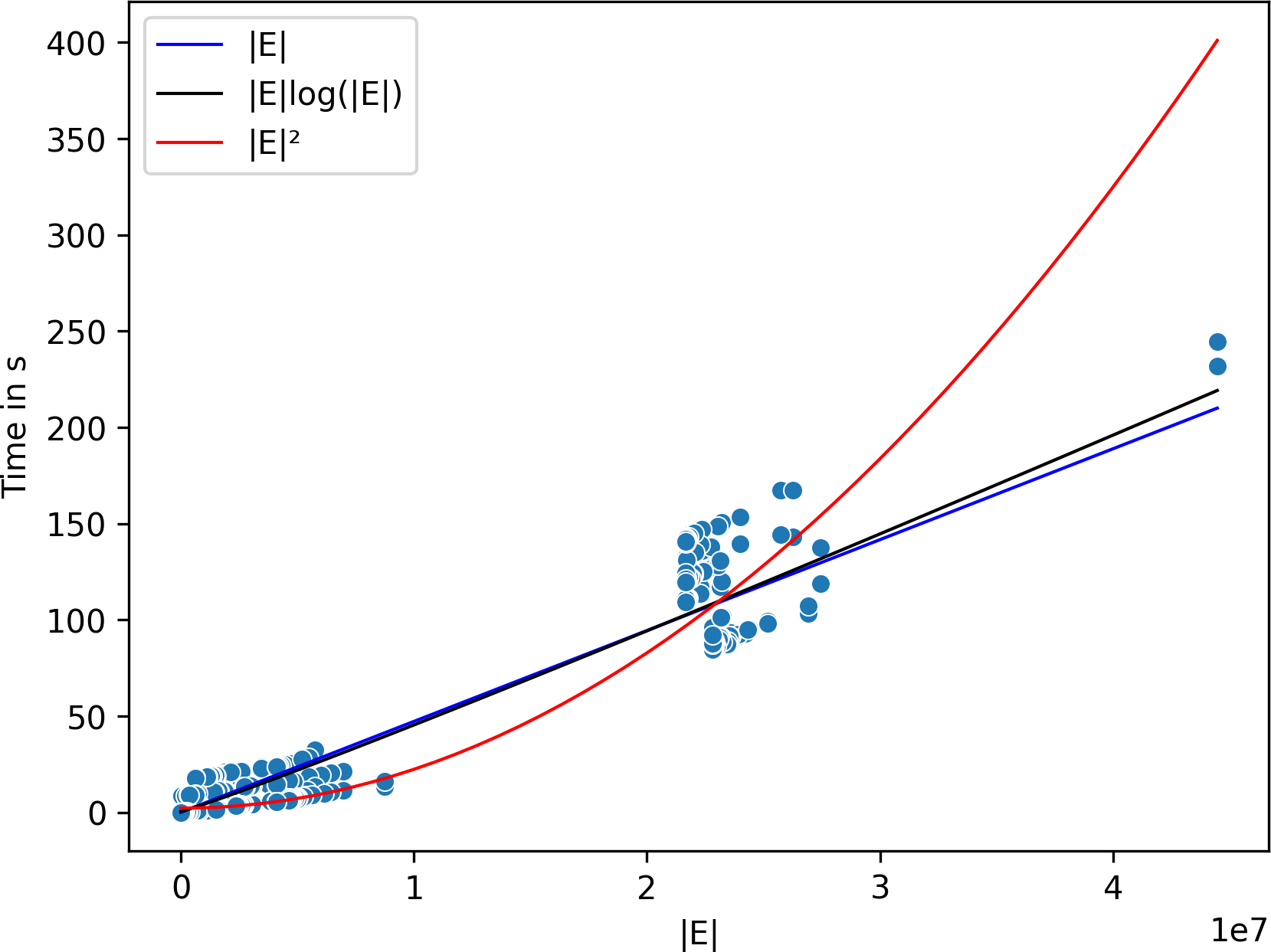}
         \caption{Model: ACC, Dataset: BTC 2019\\
         \textcolor{blue}{$R^2 = 0.96$}, 
         \textcolor{black}{$R^2 = 0.96$}, 
         \textcolor{red}{$R^2 = 0.95$ }}\label{fig:RegBTC-ACC}
     \end{subfigure}  

    \begin{subfigure}[b]{0.3\textwidth}
         \centering
         \includegraphics[width=\textwidth]{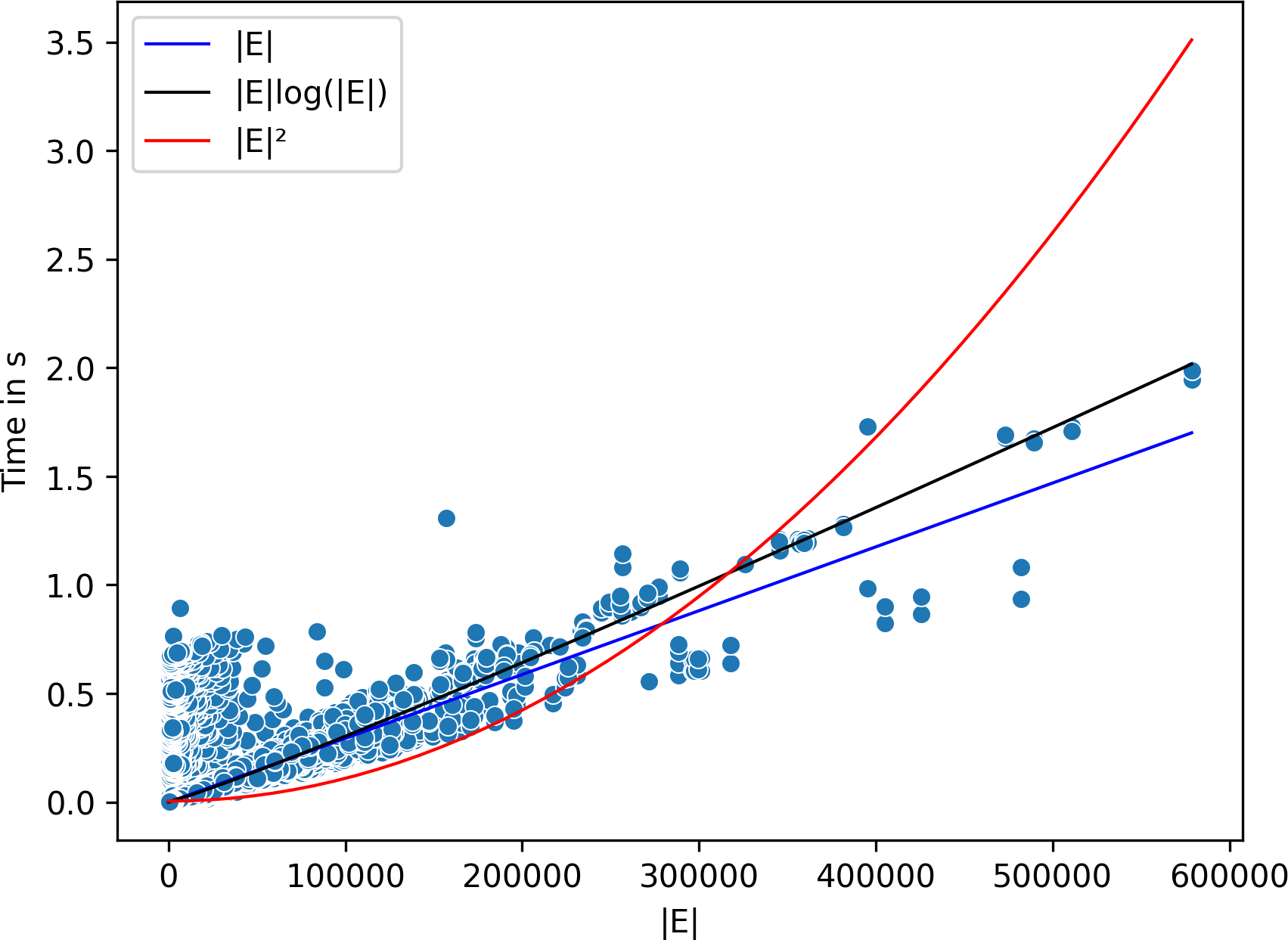}
         \caption{Model: AC, Dataset: CSS3V\\
         \textcolor{blue}{$R^2 = 0.77$}, 
         \textcolor{black}{$R^2 = 0.77$}, 
         \textcolor{red}{$R^2 = 0.54$ }}\label{fig:RegCode-AC}
     \end{subfigure}  \hfill
     \begin{subfigure}[b]{0.3\textwidth}
         \centering
         \includegraphics[width=\textwidth]{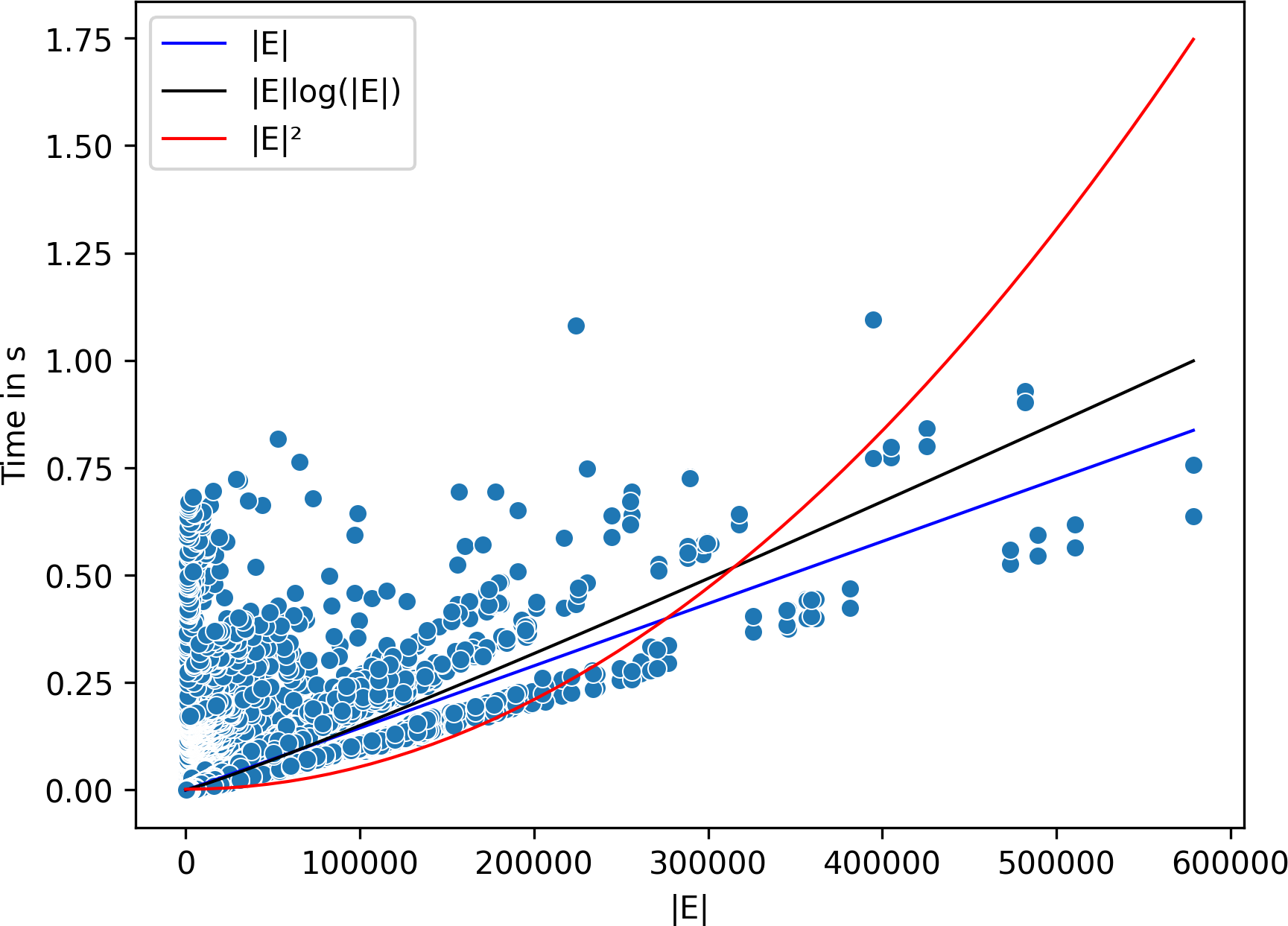}
         \caption{Model: CC, Dataset: CSS3V\\
         \textcolor{blue}{$R^2 = 0.72$}, 
         \textcolor{black}{$R^2 = 0.72$}, 
         \textcolor{red}{$R^2 = 0.50$ }}\label{fig:RegCode-CC}
     \end{subfigure}  \hfill
     \begin{subfigure}[b]{0.3\textwidth}
         \centering
         \includegraphics[width=\textwidth]{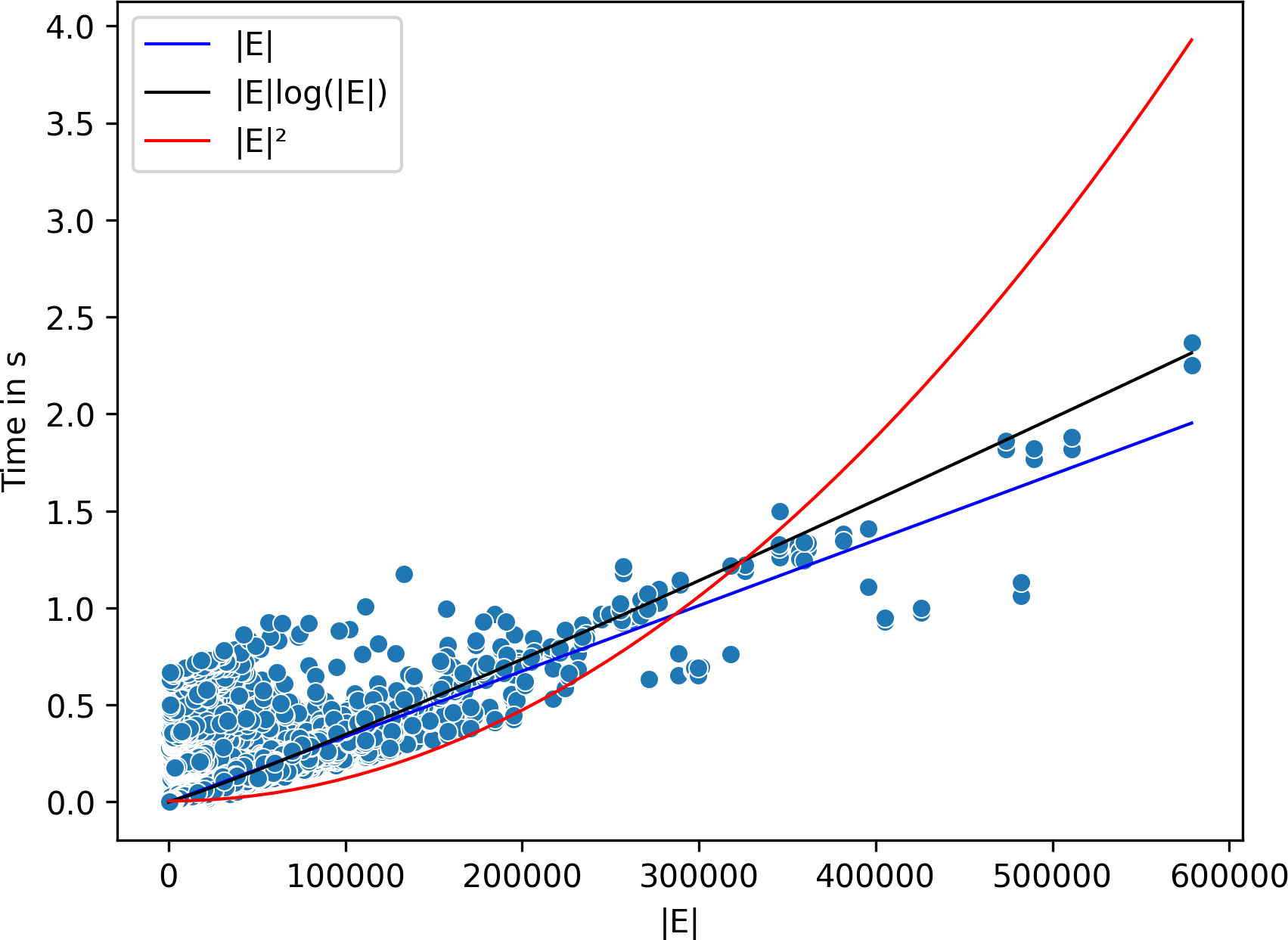}
         \caption{Model: ACC, Dataset: CSS3V\\
         \textcolor{blue}{$R^2 = 0.87$}, 
         \textcolor{black}{ $R^2 = 0.86$}, 
         \textcolor{red}{$R^2 = 0.59$ }}\label{fig:RegCode-ACC}
     \end{subfigure} 

         \begin{subfigure}[b]{0.3\textwidth}
         \centering
         \includegraphics[width=\textwidth]{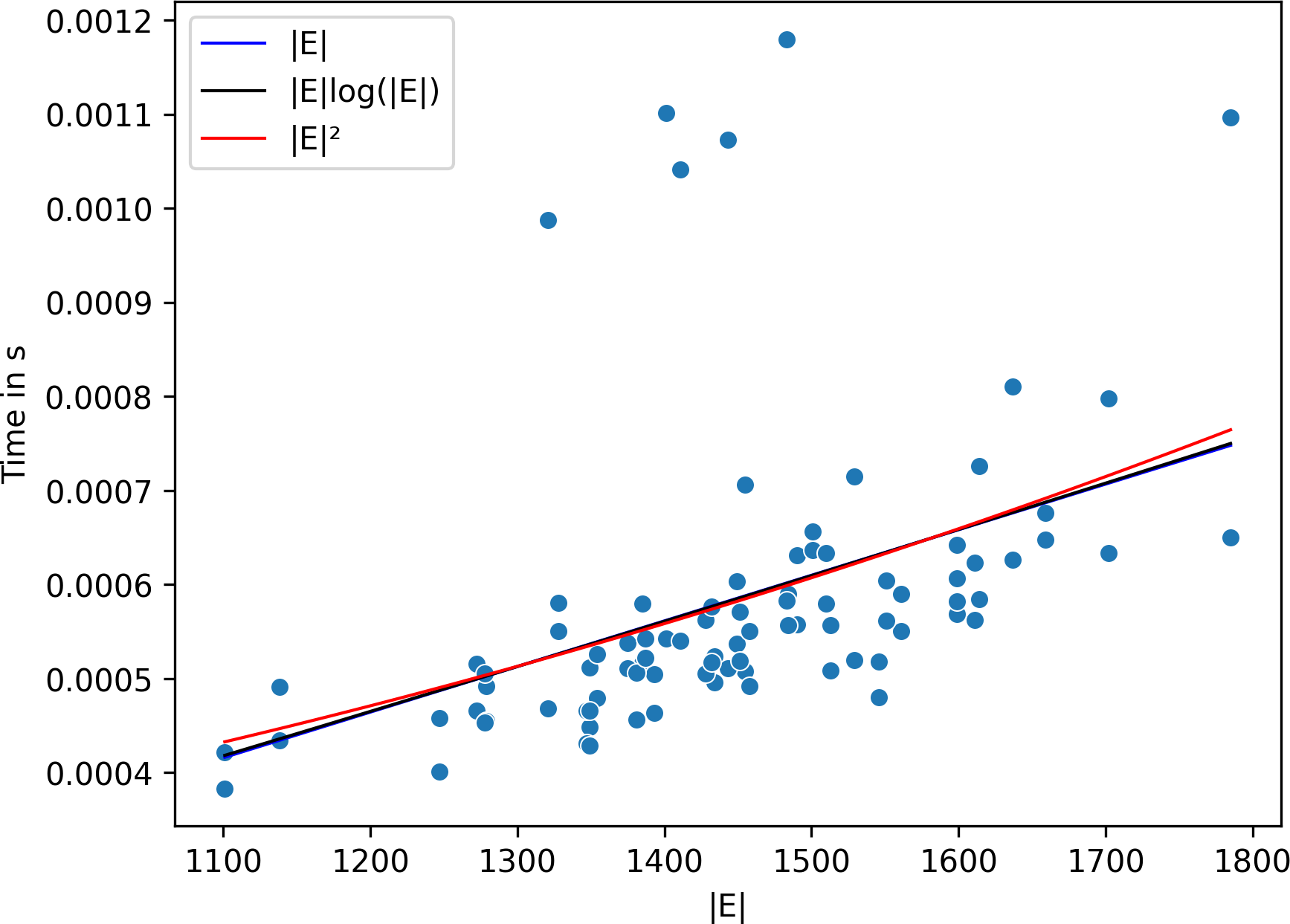}
         \caption{Model: AC, Dataset: INC 2023\\
         \textcolor{blue}{$R^2 = 0.43$}, 
         \textcolor{black}{$R^2 = 0.43$}, 
         \textcolor{red}{$R^2 = 0.43$ }}\label{fig:RegNews-AC}
     \end{subfigure}  \hfill
     \begin{subfigure}[b]{0.3\textwidth}
         \centering
         \includegraphics[width=\textwidth]{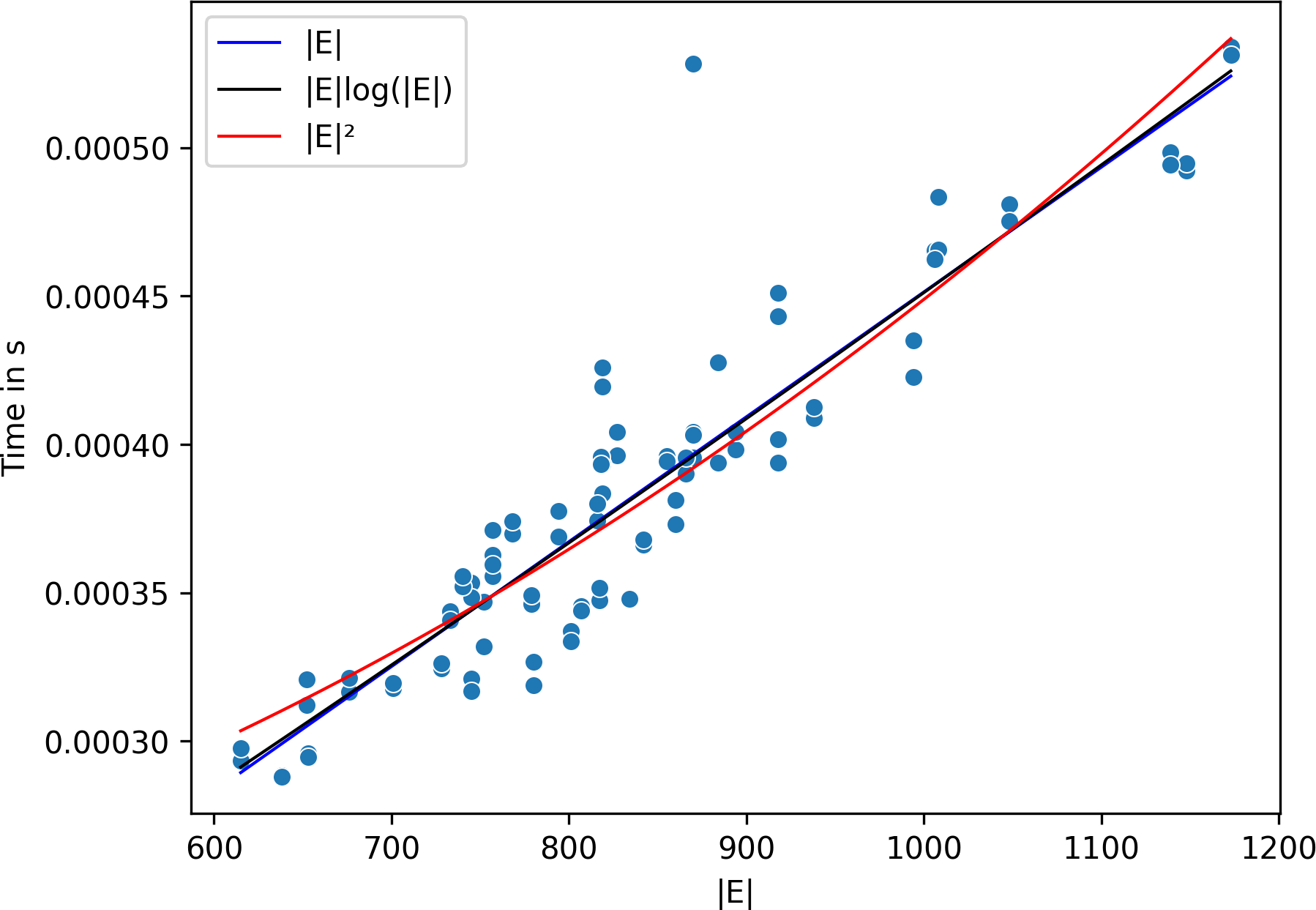}
         \caption{Model: CC, Dataset: INC 2023\\
         \textcolor{blue}{$R^2 = 0.92$}, 
         \textcolor{black}{$R^2 = 0.92$}, 
         \textcolor{red}{$R^2 = 0.91$ }}\label{fig:RegNews-CC}
     \end{subfigure}  \hfill
     \begin{subfigure}[b]{0.3\textwidth}
         \centering
         \includegraphics[width=\textwidth]{regression/NewsCCRegression.png}
         \caption{Model: ACC, Dataset: INC 2023\\
         \textcolor{blue}{$R^2 = 0.46$}, 
         \textcolor{black}{$R^2 = 0.46$}, 
         \textcolor{red}{$R^2 = 0.45$ }}\label{fig:RegNews-ACC}
     \end{subfigure} 
     \caption{Regression on the various summaries and the datasets. The row indicates the dataset, and the column indicates the summary model. For each regression, the caption contains the function and the coefficient of determination $R^2$ of $|E|$, $|E|\log (|E|)$, and $|E|^2$. The color shows to which line in the plot it belongs. All \textit{p}-values are lower than 0.05. }
    \label{fig:Regreesion}
    
\end{figure}

\end{document}